\documentclass[%
 aip,
 amsmath,amssymb,
 reprint,%
]{revtex4-1}

\usepackage{graphicx}% Include figure files
\usepackage{dcolumn}% Align table columns on decimal point
\usepackage{bm}% bold math
\usepackage{color}
\usepackage[utf8]{inputenc}
\usepackage[T1]{fontenc}
\usepackage{mathptmx}
\usepackage{etoolbox}

\makeatletter
\def\@email#1#2{%
 \endgroup
 \patchcmd{\titleblock@produce}
  {\frontmatter@RRAPformat}
  {\frontmatter@RRAPformat{\produce@RRAP{*#1\href{mailto:#2}{#2}}}\frontmatter@RRAPformat}
  {}{}
}%
\makeatother
\begin{document}

\preprint{AIP/123-QED}

\title[]{High entropy metallic glasses, what does it mean?}
% Force line breaks with \\
\author{G.V. Afonin}

\affiliation{Department of General Physics, Voronezh State Pedagogical
University,  Voronezh 394043, Russia%\\This line break forced with \textbackslash\textbackslash
}%

\author{J.C. Qiao}
%\homepage{http://www.Second.institution.edu/~Charlie.Author.}
\affiliation{School of Mechanics and Civil Architecture, Northwestern Polytechnical University, Xi’an 710072, China}%
%Second institution and/or address%\\This line break forced% with \\

\author{A.S. Makarov}
\affiliation{Department of General Physics, Voronezh State Pedagogical
University,  Voronezh 394043, Russia%\\This line break forced with \textbackslash\textbackslash
}%
\author{R.A. Konchakov}
\affiliation{Department of General Physics, Voronezh State Pedagogical
University,  Voronezh 394043, Russia}
\author{E.V. Goncharova}
\affiliation{Department of General Physics, Voronezh State Pedagogical
University,  Voronezh 394043, Russia%\\This line break forced with \textbackslash\textbackslash
}%
\author{N.P. Kobelev}
\affiliation{Institute of Solid State Physics, Russian Academy of Sciences, Chernogolovka, Moscow district 142432, Russia}
\author{V.A. Khonik*}
\email[corresponding author, email:]{ v.a.khonik@yandex.ru}

\affiliation{Department of General Physics, Voronezh State Pedagogical
University,  Voronezh 394043, Russia%\\This line break forced with \textbackslash\textbackslash
}%

\date{\today}% It is always \today, today,

\begin{abstract}
We performed calorimetric measurements on 30 bulk metallic glasses differing with their  mixing entropies $\Delta S_{mix}$. On this basis, the excess entropies $\Delta S$ and excess enthalpies $\Delta H$ of glasses with respect to their maternal crystalline states are calculated. It is found that the excess entropy $\Delta S$ on the average decreases with increasing mixing entropy $\Delta S_{mix}$. This means that so-called "high-entropy metallic glasses" (i.e. the glasses having \textit{high} $\Delta S_{mix}$) actually constitute  glasses with  \textit{low} excess entropy $\Delta S$. We predict that such glasses should have reduced relaxation ability. We also found that the excess enthalpy $\Delta H$ of glass linearly increases with its excess entropy $\Delta S$, in line with a general thermodynamic estimate.

\end{abstract}

\maketitle

Entropy-based approaches  currently constitute a new major way to the understanding of structure and physical properties of crystalline and non-crystalline metallic and non-metallic materials \cite{Murty2014,GaoSpringer2016,NemilovEntropy2018,ZhangSpringer2019,FengFundRes2022}. In particular, this applies  for  crystalline alloys, which were discovered in the beginning of the 2000s and contain five or more metallic elements each having the atomic percentage between 5 and 35 \%. These alloys can be characterized by the entropy of mixing, which reflects the changes of the entropy upon alloy preparation  and determined as 
\begin{equation}
\Delta S_{mix}=-R\sum\limits_{i=1}^nc_ilnc_i, \label{Smix}
\end{equation}
where $R$ is the universal gas constant, $c_i$ is the molar fraction of the \textit{i}-th element in the alloy and $n$ is the number of constituent elements \cite{YehAdvEngMater2004,CantorMaterSciEng2004}. The alloys with the mixing entropy $S_{mix}\geq 1.5\;R$ were named as high-entropy (HE) alloys \cite{GeorgeNatRevMater2019,DuEncMater2022,OhashiJALCOM2022,LuanJMaterSciTechnol2023}. 

A few years later it was shown that HE  melts can be solidified in  the non-crystalline state forming high-entropy metallic glasses (HEMGs)\cite{TakeuchiIntermetallics2011,GaoJNCS2011,WangJOM2014}. It was found that HEMGs have relatively high glass-forming ability \cite{WadaMaterialia2019, OhashiJALCOM2022,WangJNCS2022} and enhanced thermal stability \cite{CaoIntermetallics2018,DehkordiJNCS2022,HuoChinPhysLett2022,LuanNatComm2022}, which results in higher activation energies  \cite{ChenIntermetallics2023}. Other peculiarities of HEMGs are related to lower atomic mobility   \cite{ChenIntermetallics2023}, sluggish diffusion \cite{DuanPRL2022,JiangNatComm2021} and crystallization kinetics \cite{JiangJA2016,YangMaterResLett2018}, slow dynamics of homogeneous flow \cite{ZhangScrMater2022},
  decreased dynamic and spatial heterogeneities  \cite{JiangNatComm2021}. HEMGs were found to have good  mechanical properties \cite{HuoChinPhysLett2022,LuanJMaterSciTechnol2023,QiIntermetallics2015} and sometimes display superior magnetic characteristics \cite{QiIntermetallics2015,XuJNCS2018,ChenJALCOM2021}, superplastic behavior above the glass transition temperature $T_g$ \cite{DehkordiJNCS2022,ZhangIntJPlast2022}, unique biomedical properties \cite{WangJOM2014} as well as excellent irradiation tolerance \cite{WangJNucLMater2019}. It is sometimes said that HEMGs combine the features of both metallic glasses and high entropy crystalline alloys  \cite{ChenJALCOM2021}.

Like HE crystalline alloys, HEMGs are defined as glasses having high mixing entropy $\Delta S_{mix}/R\geq 1.5$  defined by Eq.(\ref{Smix}). However, despite of intense investigations in the past decade, it is still unclear what does high $\Delta S_{mix}$ mean from a physical point and to what consequences does it lead. This seems to be a major issue of the physics of high-entropy non-crystalline materials. This work is addressed to this problem. We determined the excess entropy $\Delta S$ of 30 bulk metallic glasses with respect to their maternal crystalline states and compared it with the mixing entropy $\Delta S_{mix}$ accepted as a state variable. We found that on the average the \textit{high mixing entropy} $\Delta S_{mix}$ corresponds to the \textit{low excess entropy} $\Delta S$. This conclusion appears to be quite expectable and understable and explains why  HEMGs  should have reduced relaxation ability. Besides that, we show that an increase of  the excess entropy leads to a linear growth of the excess enthalpy of glass, which is also quite explainable. 

The determination of MGs' entropy (as well as other thermodynamic potentials) was considered long ago (e.g. Ref.\cite{BuschJApplPhys1998}) and the works basing on the thermodynamic approach have continued to the present   \cite{SchaweThermochimActa2020,NeuberActamater2021}. Such works utilize heat capacity data in the liquid state and/or below $T_g$ together with the heat of fusion for the calculation of the excess thermodynamic potentials of the supercooled liquid state with respect to the maternal crystal. The obtained results were used mainly for an estimate of the glass-forming ability of supercooled liquids and their crystallization kinetics.  

Recently, we suggested a method to calculate the excess entropy of \textit{solid} glass $\Delta S$ with respect to the maternal crystal (by this crystal we understand a polycrystalline structure that arises as a result of the complete crystallization of glass and does not undergo any subsequent phase transformations), which is defined in line with classical thermodynamics using calorimetric data as \cite{MakarovJPCM2021b,MakarovJETPLett2022}
\begin{equation}
\Delta S(T)=\frac{1}{\dot{T}}\int_{T}^{T_{cr}} \frac{\Delta W(T)}{T}dT, \label{DeltaS}
\end{equation}
where $\Delta W$ is the differential heat flow specified below, $T_{cr}$ is the temperature of the complete crystallization and $\dot{T}$ is the heating rate. To calculate temperature dependence $\Delta S(T)$ with this equation, we performed differential scanning calorimetry (DSC) measurements using a Hitachi DSC 7020 instrument operating in high-purity (99.999\%) $N_2$ atmosphere at a rate of 3 K/min. For this investigation, we tested 30  X-ray amorphous bulk metallic glasses (MGs) prepared by melt suction. The compositions of these MGs are listed  in Table 1 in the order with their mixing entropy $\Delta S_{mix}/R$ calculated using Eq.(\ref{Smix}). It is seen that the mixing entropy changes in the range $0.83\leq \Delta S_{mix}/R \leq 1.79$, i.e. covers almost the whole possible range, from a minimal value $\Delta S_{mix}/R=0.83$ for a three-component MG to $\Delta S_{mix}/R=1.79$ for a HE equiatomic six-component glass. 

To decrease the effect of the preparation conditions on the excess entropy, all glasses  were tested in the relaxed (preannealed) state according to the following protocol: \textit{i}) initial sample was heated up to the temperature of the complete crystallization $T_{cr}$ with empty reference DSC cell. This crystallized sample was next moved to the reference DSC cell; \textit{ii}) a new sample in the initial state was heated into the supercooled liquid state (i.e. between $T_g$ and the crystallization onset temperature) and cooled back to room temperature at the same rate of 3 K/min;  \textit{iii)} the same (relaxed) sample was tested up to $T_{cr}$ (run 2) and, finally, \textit{iv)} the same (crystallized) specimen is again heated up to $T_{cr}$ (run 3). This protocol allows performing  DSC measurements with the reference cell containing fully crystallized sample of approximately the same mass ($\approx$  70 mg) so that the measured heat flow $\Delta W$ constitutes the difference between the heat flow coming from relaxed glass $W_{gl}$ and its crystalline counterpart $W_{cr}$ and termed  as the differential heat flow, i.e. $\Delta W=W_{gl}-W_{cr}$, which enters Eq.(\ref{DeltaS}). It is seen that if temperature $T=T_{cr}$ then $\Delta S=0$ and, therefore, the quantity $\Delta S$ given by this equation represents the excess entropy of solid glass with respect to the maternal crystal.

\begin{figure}[t]
\center{\includegraphics[scale=0.65]{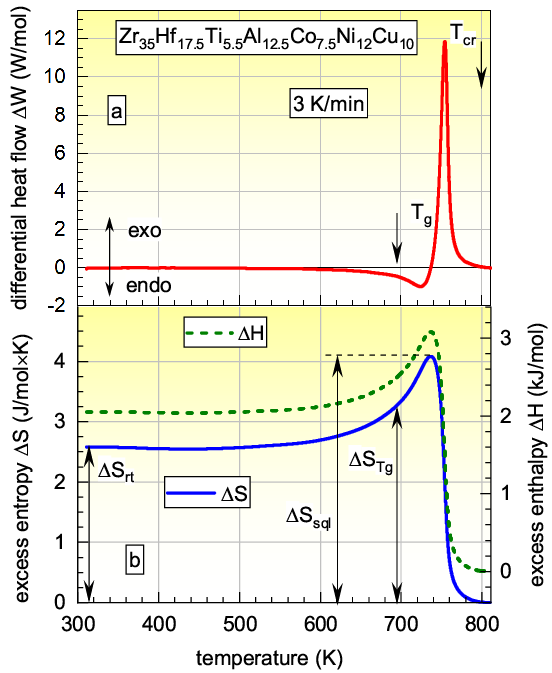}}% Here is how to import EPS art
\caption[*]{\label{Fig1.eps} Differential heat flow of relaxed HE glassy Zr$_{35}$Hf$_{17.5}$Ti$_{5.5}$Al$_{12.5}$Co$_{7.5}$Ni$_{12}$Cu$_{10}$ (a) and temperature dependences of its excess entropy $\Delta S$ and excess enthalpy $\Delta H$ calculated with Eqs (\ref{DeltaS}) and (\ref{DeltaH}), respectively (b). The glass transition temperature $T_g$ and the temperature of the complete crystallization $T_{cr}$ are shown by the arrows. The excess entropies at room temperature ($\Delta S_{rt}$), glass transition temperature ($\Delta S_{Tg}$) and in the supercooled liquid state ($\Delta S_{sql}$) are indicated.}  
\end{figure}

Panel (a) in Fig.\ref{Fig1.eps} gives  the differential heat flow $\Delta W$ for a relaxed HE glass Zr$_{35}$Hf$_{17.5}$Ti$_{5.5}$Al$_{12.5}$Co$_{7.5}$Ni$_{12}$Cu$_{10}$ taken as an example. The plot illustrates common regularities of DSC testing on preannealed MGs given by an endothermal effect below $T_g$ and in the supercooled liquid state (above $T_g$) and strong crystallization-induced exothermal reaction at higher temperatures up to the temperature of the complete crystallization $T_{cr}$. Panel (b) shows the excess entropy $\Delta S$ calculated with Eq.(\ref{DeltaS}) using $\Delta W$-data given in panel (a). The excess entropy  is first nearly constant (up to $\approx 500$ K), then increases with temperature due to the endothermal heat flow below and near $T_g$, next rapidly increases due to strong endothermal disordering in the supercooled liquid state and finally falls  to zero because of the complete crystallization at $T=T_{cr}$. This is a typical $\Delta S(T)$-pattern generally characteristic of all MGs under investigation. The characteristic values of the excess entropy at room temperature ($\Delta S_{rt}$), at the glass transition temperature ($\Delta S_{Tg}$) and in the supercooled liquid state ($\Delta S_{sql}$) are indicated.

\begin{table}[t]
%\begin{center}
%\footnotesize
\caption{\label{tab:table1} Metallic glasses under investigation and their mixing entropies $\Delta S_{mix}/R$ calculated with Eq.(\ref{Smix}). } 

\scriptsize
\begin{tabular}{p{3mm}|c|c|c|c|c|c|c|p{6mm}|c}
\hline
\hline
No & Composition (at.\%)&$\Delta S_{mix}/R$&No&Composition (at.\%)&$\Delta S_{mix}/R$ \\ 
\hline
\hline
%Zr47.5Cu47.5Al5

 1 & Cu$_{48}$Zr$_{48}$Al$_4$ & 0.83 &16 &Zr$_{48}$Cu$_{34}$Ag$_{8}$Al$_{8}$Pd$_2$& 1.20\\

2 & Zr$_{47.5}$Cu$_{47.5}$Al$_{5}$  & 0.86 & 17 &Pd$_{43.2}$Cu$_{28}$Ni$_{8.8}$P$_{20}$ & 1.25  \\ 

3 & Zr$_{46}$Cu$_{46}$Al$_{8}$ & 0.92 & 18 &Zr$_{57}$Nb$_{5}$Al$_{10}$Cu$_{15.4}$Ni$_{12.6}$&1.25\\

4 & Cu$_{49}$Hf$_{42}$Al$_{9}$ & 0.93  & 19 &Pt$_{42.5}$Cu$_{27}$Ni$_{9.5}$P$_{21}$&1.26\\

5 & La$_{55}$Ni$_{10}$Al$_{35}$ & 0.93 & 20 &Pd$_{40}$Cu$_{30}$Ni$_{10}$P$_{20}$&1.28\\

6 & Zr$_{50}$Cu$_{40}$Al$_{10}$& 0.94 &21 &Zr$_{52.5}$Ti$_{5}$Cu$_{17.9}$Ni$_{14.6}$Al$_{10}$&1.31\\

7 & Zr$_{47}$Cu$_{45}$Al$_{7}$Fe$_1$& 0.95 &22 &Zr$_{31.6}$Hf$_{13.4}$Al$_{8.7}$Ag$_{8.4}$Cu$_{37.8}$&1.42 \\

8 & Zr$_{56}$Co$_{28}$Al$_{16}$& 0.97 &23 &Ti$_{32.8}$Zr$_{30.2}$Cu$_{9}$Ni$_{5.3}$Be$_{22.7}$&1.45\\

9 & Zr$_{46}$Cu$_{45}$Al$_7$Ti$_2$& 0.98 &24 &Zr$_{31}$Ti$_{27}$Be$_{26}$Cu$_{10}$Ni$_{6}$&1.46 \\

10 & La$_{55}$Al$_{25}$Co$_{20}$& 0.99 &25  &Ti$_{20}$Zr$_{20}$Hf$_{20}$Cu$_{20}$Be$_{20}$&1.61\\

11 & Zr$_{55}$Co$_{25}$Al$_{20}$& 0.99 &26 &Ti$_{20}$Zr$_{20}$Hf$_{20}$Ni$_{20}$Be$_{20}$&1.61\\

12 & Zr$_{65}$Al$_{10}$Ni$_{10}$Cu$_{15}$& 1.03 &27 &Zr$_{35}$Hf$_{13}$Al$_{11}$Ag$_{8}$Ni$_{8}$Cu$_{25}$&1.63  \\

13 &Pd$_{40}$Ni$_{40}$P$_{20}$& 1.05 &28  &Zr$_{40}$Hf$_{10}$Ti$_{4}$Y$_{1}$Al$_{10}$Cu$_{25}$Ni$_{7}$Co$_{2}$Fe$_{1}$&1.66\\

14 &Zr$_{50}$Cu$_{34}$Ag$_{8}$Al$_8$ & 1.12 &29 & Zr$_{35}$Hf$_{17.5}$Ti$_{5.5}$Al$_{12.5}$Co$_{7.5}$Ni$_{12}$Cu$_{10}$&1.77 \\

15 &Zr$_{46}$Cu$_{36.8}$Ag$_{9.2}$Al$_8$ & 1.15 &30 & Ti$_{16.7}$Zr$_{16.7}$Hf$_{16.7}$Cu$_{16.7}$Ni$_{16.7}$Be$_{16.7}$&1.79 \\
 
\hline
\hline

\end{tabular}
%\end{ruledtabular}
%\end{center}
\end{table}

\begin{figure}[t]
\center{\includegraphics[scale=0.65]{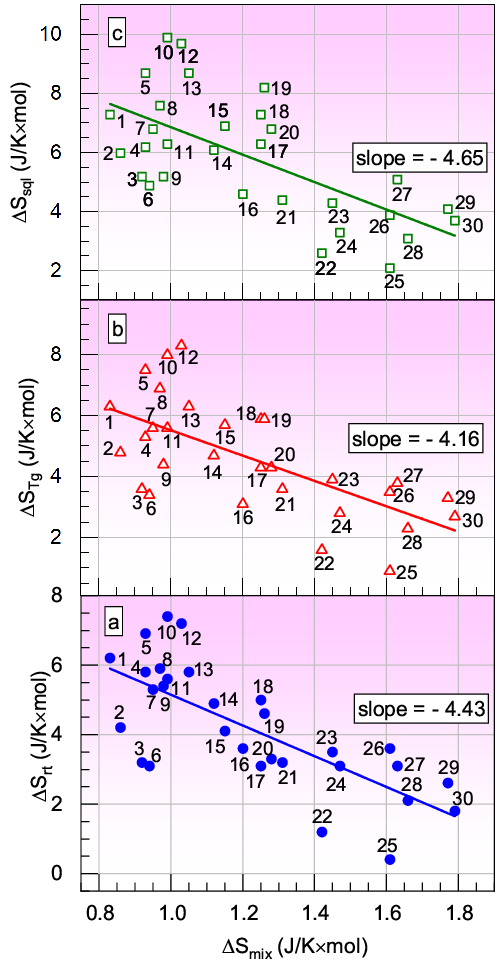}}% Here is how to import EPS art
\caption[*]{\label{Fig2.eps} Excess entropy at room temperature $\Delta S_{rt}$ (a), at the glass transition temperature $\Delta S_{Tg}$ (b) and in the supercooled liquid state $\Delta S_{sql}$ (c) as function of the mixing entropy $\Delta S_{mix}$. The numbers indicate glass compositions according to Table 1. The straight lines give least-square-fit approximations. The corresponding slopes $d\Delta S_{rt}/d\Delta S_{mix}$, $d\Delta S_{Tg}/d\Delta S_{mix}$ and $d\Delta S_{sql}/d\Delta S_{mix}$ in panels (a), (b) and (c) are indicated. It is seen that the excess entropy $\Delta S$ in any case  on the average decreases with the mixing entropy $\Delta S_{mix}$. }  
\end{figure}

Figure \ref{Fig2.eps} gives the dependences of the excess entropy $\Delta S_{rt}$ (panel (a)), $\Delta S_{Tg}$ (panel (b)) and $\Delta S_{sql}$ (panel (c)) as a function of the mixing entropy $\Delta S_{mix}$. The numbers indicate the glass compositions according to Table 1. It is seen that, despite of some  scatter,  the excess entropies on the average decrease with the mixing entropy as evidenced  by the least square linear fits. At that, the slopes of the linear fits in panels (a) to (c) are equal within $\approx 6\%$ or even less. 

Thus, one can conclude that the excess entropy of glass with respect to the maternal crystal decreases with the mixing entropy, independent of what quantity ($\Delta S_{rt}$, $\Delta S_{Tg}$ or $\Delta S_{sql}$) is taken to quantify this dependence. In general, the origin of this dependence seems to be clear. The high mixing entropy for many-component MGs assumes that the supercooled melt is strongly relaxed (stabilized \cite{GeorgeNatRevMater2019}) due to accomodative movements of different atoms into many of available potential wells in the structure. This relaxed melt is then quenched to form a solid glass, which to a certain extent inherits  the relaxed state of the melt. Thus, the more is the mixing entropy, the more the degree of glass relaxation should be. It is also important to emphasize that the mixing entropy $\Delta S_{mix}$ does not depend on the state of glass (i.e. does not vary upon structural relaxation) and, moreover, has the same value both in the amorphous and crystalline states. Thus, the mixing entropy $\Delta S_{mix}$ is not a part  of the excess entropy $\Delta S$.

On the other hand, the excess entropy $\Delta S$ characterizes the disorder of glass  and, therefore, the level of its non-equilibrium (see  Ref.\cite{MakarovScrMater2024} for a discussion). Thus, it is clear that glass excess entropy should decrease with the mixing entropy.  At that, while $\Delta S_{mix}$ reflects only the percentage of different atoms, the excess entropy accounts for all glass parameters  (chemical composition, interatomic interaction, melt quenching rate, entropy changes due to glass formation, etc.). This seems to be the origin of the data scatter in Fig.\ref{Fig2.eps}. It is also worthy of notice  that $\Delta S$ in any case is notably bigger than $\Delta S_{mix}$ (see Fig.\ref{Fig2.eps}) pointing out a major role of the excess entropy in the glass formation, as argued earlier \cite{MakarovScrMater2024}. 

Since the excess entropy directly reflects the disorder in glass  \cite{MakarovScrMater2024}, one has also to  conclude that MGs with higher $\Delta S$ should be more prone to structural relaxation below the glass transition. Conversely, glasses with lower $\Delta S$ (i.e. higher $\Delta S_{mix}$) should be more ordered and more resistant to structural relaxation upon thermal and other external impact. A number of HEMGs' peculiarities mentioned above (enhanced thermal stability, higher activation energies,  lower atomic mobility, sluggish diffusion and crystallization kinetics, etc.) are in line with this conclusion. It is to be noted also that one-component glasses (e.g. molecular organic and inorganic glasses, with SiO$_2$ as an example) have zero mixing entropy. For such glasses the factors determining their excess entropy and relaxation ability, in addition to those mentioned above, should be revealed.

\begin{figure}[t]
\center{\includegraphics[scale=0.65]{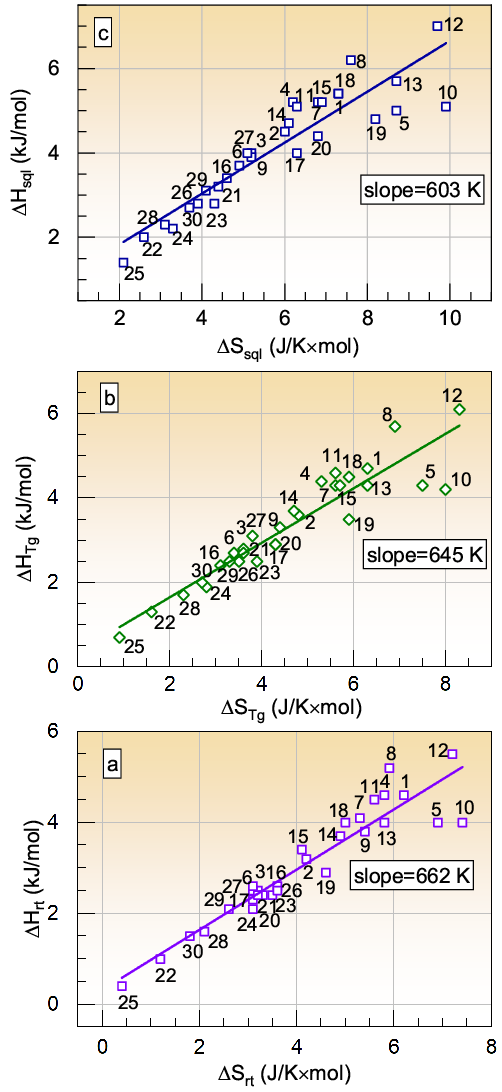}}% Here is how to import EPS art
\caption[*]{\label{Fig3.eps} Excess enthalpy at room temperature $\Delta H_{rt}$ (a), at the glass transition temperature $\Delta H_{Tg}$ (b) and in the supercooled liquid state $\Delta H_{sql}$ (c) as a function of the corresponding excess entropies $\Delta S_{rt}$, $\Delta S_{Tg}$ and $\Delta S_{sql}$. The straight lines gives least-square-fit approximations. The slopes of these lines are equal to 637 K with an error of $\approx 10\%$ or less.  }  
\end{figure} 

Calorimetric measurements performed in the present study also allow the calculation of the excess enthalpy of glass $\Delta H$ as \cite{MakarovJPCM2021b,MakarovJETPLett2022}
\begin{equation}
\Delta H(T)=\frac{1}{\dot{T}}\int_{T}^{T_{cr}} \Delta W(T)dT, \label{DeltaH}
\end{equation}
where all quantities are  defined above. Figure \ref{Fig1.eps}(b) gives an example of $\Delta H(T)$ temperature dependence, which is calculated with DSC thermogram in Fig.\ref{Fig1.eps}(a) using Eq.(\ref{DeltaH}). It is seen that temperature dependence of the excess enthalpy is similar to that of the excess entropy. The characteristic values of the excess enthalpy at room temperature $(\Delta H_{rt}$), glass transition temperature $(\Delta H_{Tg}$) and in the supercooled liquid state $(\Delta H_{sql}$) are defined similarly to the corresponding values of the excess entropy shown in Fig.\ref{Fig1.eps}(b). Panels (a), (b) and (c) in Fig.\ref{Fig3.eps} show the dependences of the excess enthalpy at room temperature $\Delta H_{rt}$, at the glass transition temperature $\Delta H_{Tg}$ and in the supercooled liquid state $\Delta H_{sql}$ as functions of the corresponding excess entropies $\Delta S_{rt}$, $\Delta S_{Tg}$ and $\Delta S_{sql}$, respectively.  It is seen that the excess enthalpies in all cases linearly increase with the excess entropies and the corresponding slopes $\Delta S/\Delta H$ are equal to 637 K with an uncertainty of $\approx 10\%$ or less. This result is quite reasonable since the enthalpy change due to glass formation at $T_g$  can be estimated using a general thermodynamic relation as $\Delta H_{Tg}\approx T_g\Delta S_{Tg}$, where $\Delta S_{Tg}$ is the corresponding entropy change. Then, the derivative  derivative $d\Delta H_{Tg}/d\Delta S_{Tg}$ should be equal to the glass transition temperature. Since the experimental value of this derivative is 637 K while the glass transition temperature averaged over all 30 MGs listed in Table 1 is 626 K, one can state that the above calculations of the excess entropy and excess enthalpy are in a good agreement with general thermodynamic consideration.  

It is to be emphasized that the excess enthalpy  $\Delta H$ and its temperature dependence  can be to a very good precision described as a result of the evolution of the system of frozen-in  defects similar to dumbbell interstitials in crystalline metals. As shown earlier, $\Delta H$ at any temperature equals to the elastic energy of these defects accurate to within 10--15\% and crystallization of glass can be considered just as the dissipation of this elastic energy into heat \cite{MakarovJPCM2021b,MakarovJETPLett2022}. Since the excess entropy $\Delta S$  is uniquely determined by the excess enthalpy $\Delta H$, the former is also intrinsically linked to glass defect system and its changes upon heat treatment. In turn, the degree of glass disorder described by $\Delta S$ (as demonstrated in Ref.\cite{MakarovScrMater2024}) is related to the defect system of glass as well.

Finally, it is to be noted that the excess entropy $\Delta S$ can be used for the calculation of the relaxation time within the framework of the Adam-Gibbs model. This consideration shows that $\Delta S$ calculated near $T_g$ is inversely proportional to the thermodynamic fragility \cite{KonchakovJETPLett2024}, which is known to be a major characteristic of supercooled liquids and glasses. Thus, the fragility is also linked to the degree of disorder and defect structure of glass as suggested earlier \cite{MakarovJPCM2021}.      

In conclusion, we performed calorimetric measurements of 30 bulk metallic glasses with the mixing entropies $\Delta S_{mix}/R$ changing from 0.8-0.9 for simple three-component MGs to 1.7--1.8 for high-entropy six-component  glasses. On this basis, we calculated the excess entropy and excess enthalpy of glass with respect to the maternal crystalline state using Eqs.(\ref{DeltaS}) and (\ref{DeltaH}), which directly follow from classical thermodynamics without any additional assumptions. First, we found that the excess entropies determined for the characterisitic temperatures (room temperature, glass transition temperature and the peak temperature of the excess entropy in the supercooled liquid state) on the average decrease with the mixing entropy. This leads to an answer formulated in the title of this Letter as: so-called "high-entropy metallic glasses" with \textit{high} mixing entropy actually constitute the glasses with relatively \textit{low} excess entropy, which are, consequently, least disordered.  We therefore predict that "high entropy metallic glasses" should display reduced relaxation ability upon thermal or other external impacts.       

We also show that the excess enthalpy $\Delta H$ at any of the above characteristic temperatures linearly increases with the excess entropy $\Delta S$ and the derivative $d\Delta H/d\Delta S$ is close to the glass transition temperature, in a good agreement with general thermodynamic estimate of the entropy and enthalpy changes upon glass transition.     

The obtained results bring an important question on how the high-excess-entropy state is related to mechanical and other physical behaviors of metallic glasses.  The corresponding studies are on the agenda.

\section*{AUTHOR DECLARATIONS}

\section*{Conflict of Interest}
The authors have no conflicts to disclose.

\section*{Author Contributions}
G.V. Afonin: experimental investigation and data analysis; J.C. Qiao: samples, 
discussion, editing; A.S. Makarov: experimental investigation and data analysis; 
 R.A. Konchakov: data analysis; E.V. Goncharova: data analysis; N.P. Kobelev: discussion and editing; V.A. Khonik: general conceptualization, manuscript preparation.

\section*{DATA AVAILABILITY}
The data that support the findings of this study are available from
the corresponding authors upon reasonable request.

\section*{Acknowledgments}

This work was supported by the Russian Science Foundation under the grant No 23-12-00162. 

%%%%%%%%%%%%%%%%%%%%%%%%%%%%%%%%%%%%%%%%%%
\vspace{6pt}

%\authorcontributions{The authors equally contributed to this work.}

\end{document}